\pdfoutput=1
\documentclass[11pt]{article}

\usepackage[]{acl}
\usepackage{amsmath}

\usepackage{times}
\usepackage{latexsym}
 \usepackage{graphicx}
  \usepackage{multirow}
  
  \usepackage{multicol}
\usepackage{adjustbox}
 
\usepackage[T1]{fontenc}
\usepackage[utf8]{inputenc}

\usepackage{microtype}
\usepackage{amsfonts}
\title{CIRCLE: Multi-Turn Query Clarifications with Reinforcement Learning}

\author{Pierre Erbacher \\
  Sorbonne Université \\
  Paris, France \\
  \small\texttt{pierre.erbacher@isir.upmc.fr} \\\And
  Laure Soulier \\
  Sorbonne Université \\
  Paris, France \\
  \small\texttt{laure.soulier@isir.upmc.fr} \\}

\begin{document}
\maketitle
\begin{abstract}
Users often have trouble formulating their information needs into words on the first try when searching online. This can lead to frustration, as they may have to reformulate their queries when retrieved information is not relevant. This can be due to a lack of familiarity with the specific terminology related to their search topic, or because queries are ambiguous and related to multiple topics. Most modern search engines have interactive features that suggest clarifications or similar queries based on what others have searched for. However, the proposed models are either based on a single interaction or evaluated on search logs, hindering the naturalness of the interactions. 
In this paper,  we  introduce CIRCLE, a  generative model for multi-turn query  Clarifications wIth ReinforCement LEarning that leverages multi-turn interactions through a user simulation framework.  Our model  aims at generating a diverse set of query clarifications using a pretrained language model fine-tuned using reinforcement learning. We evaluate it against well established google suggestions using a user simulation framework.
\end{abstract}

% these suggestion features have some limitations. First, they rely on search engine logs that have been recorded on a mass scale, which are not always available. Second, they propose only one interaction, not assisting users thought multiple interaction steps. Multiple studies have shown the benefit of multi-turn clarification to assist the user and provide additional information to the system in order to improve the accuracy of the results.

\section{Introduction}

The recent advances in deep learning have fostered the Information retrieval (IR) community to design  more effective  ranking models \cite{Khattab-Colbert,Hofstatter-topicaware}. Although they rely on the notion of semantics aiming at bridging the gap between the query and the document vocabularies, the formulation  of information needs into expressive and complete queries remains a critical step for solving IR tasks. This is particularly the case when the information need is complex or multi-faceted, or either when the resulting query is ambiguous or vague. In parallel to the cognitive aspect of the information modeling, one of the reason from the user side might be the lack of knowledge and vocabulary associated to the searched domain. \cite{DBLP:conf/trec/Bhavhani01,10.1145/1390334.1390506,10.1002/asi.10367} show that users searching information outside their expertise domain use under specified queries, leading to less effective search results. Even when users are seeking information in  previously experienced search topics, users may still not be able to express intent using the right vocabulary. Arguello et al. \cite{10.1145/3406522.3446021} define the Tip of the Tongue Phenomenon (TOT) as the cognitive state where users are unable to retrieve known items from memory, for example names or words associated to the search intent, also leading to less effective queries. %The attention of the IR community is drawn to these causes because they are inherent to users and explain the need for query refinement and intent clarification.

To tackle this issue, the IR community has focused its attention  on the understanding of users' information needs, starting with post-hoc reformulation methods and sliding more and more towards proactive interactive approaches \cite{10.1145/3397271.3401206}. 
Early approaches have investigated query reformulation aiming at rewriting the query to improve the results returned by a search engine by using more appropriate words \cite{Amati-reformulation,Lavenkro-reformulation,Rocchio-SMART-1971,zukerman-raskutti-2002-lexical,DBLP:journals/corr/abs-1809-10658,10.1145/1835449.1835546,DBLP:journals/corr/abs-1303-0667,10.1145/2851613.2851696,10.1145/3130348.3130376}.
%One first line of work relies on query reformulation aiming at rewriting the query to improve the results returned by a search engine by using more appropriate words \cite{Amati-reformulation,Lavenkro-reformulation,Rocchio-SMART-1971,zukerman-raskutti-2002-lexical, LEARNING TO COORDINATE MULTIPLE REINFORCEMENT LEARNING AGENTS FOR DIVERSE QUERY REFORMULATION, [RELEVANCE MODEL (RM3)expansion (Lavrenko & Croft, 2001).]} 
 % A lot of effort has been provided by designing models based on either (pseudo-)relevance feedback \cite{Amati-reformulation,Lavenkro-reformulation,Rocchio-SMART-1971} or external knowledge resources \cite{zukerman-raskutti-2002-lexical}. Recently, the advances in machine translation models and in large language models have turned this task into a query generation task \cite{Dalton-CAST,elgohary-etal-2019-unpack}.  
 However, these methods assume the most probable topic for a query and does not always resolve the ambiguity.
 Other strategies have been proposed in the reformulation process such as the search/query diversification \cite{Agrawal-diversity,Fei2016-diversification, 10.1145/290941.291025,DBLP:journals/corr/abs-2108-04026,48553,10.1145/1772690.1772780,dang-diversity} to increase the query coverage or the leveraging of search history to infer user's profile or session context \cite{Matthijs-UserHistory, Weize-PreSearchContext,Harvey2013BuildingUP, Xiang2010ContextawareRI, Bennett2012ModelingTI}. 
 But these approaches might first frustrate the user with the release of top-ranked documents not always relevant \cite{Wang-limitdiversification} or be hindered with the user's behavior's variability over time \cite{Analyzing-clarification}.

Rather than reformulating queries or adapting ranking models to the user, a promising approach is to include the user in the clarification process by interacting with him/her thought clarifying questions or query suggestions \cite{Aliannejadi19-askingclarifying,erbacher2022, Zamani-Clarification,Wu2018QuerySW,Guo2011IntentawareQS,Santos2012LearningTR}. While a lot of progress as been made in this area, proposed work have multiple limitations. They are limited to a single interaction turn \cite{Zamani-Clarification,Wu2018QuerySW,Guo2011IntentawareQS,Santos2012LearningTR}. \cite{Aliannejadi19-askingclarifying} show the benefit of multi-turn clarification, however the method relies on limited crowd-sourcing and predefined sequences of interactions for specific topics. Very recently, \cite{erbacher2022} has proposed a simulation framework enabling to model the interactions between the user and an IR system aiming at clarifying questions and retrieving document rankings.

In this paper, we introduce  a multi-turn query clarification model aiming at generating new and diverse query suggestions between each interaction turn. Our model relies on pre-trained language models and leverages both supervised learning and reinforcement learning to clarify the initial query issued by the user.  More particularly, we fine-tune the language model with a policy at the sequence generation level for predicting the next token in the sequence such as in \cite{Ouyang2022TrainingLM}. In addition, our policy contributes to the  generation of a set of clarification queries which are optimized to balance diversity and effectiveness. 
Based on \cite{erbacher2022},  we integrate our multi-turn query clarification model into their proposed simulation framework to mimic user-system interactions. 
It is worth noting that in such framework we expect our model to embed a multi-turn level policy that optimizes the trajectory over all interactions. We let this perspective for future work since it will imply a very sparse reward, and training such policy is more complex. 

To show the benefit of multi-turn interactivity and evaluate the effectiveness of our model, we conduct an experimental evaluation using the MS Marco dataset in a simulated framework.  We compare our results against well-established search engines that can rely on logs of billions of users to construct query suggestions.

\begin{figure*}[t]
    \centering
    \includegraphics[width=0.85\textwidth]{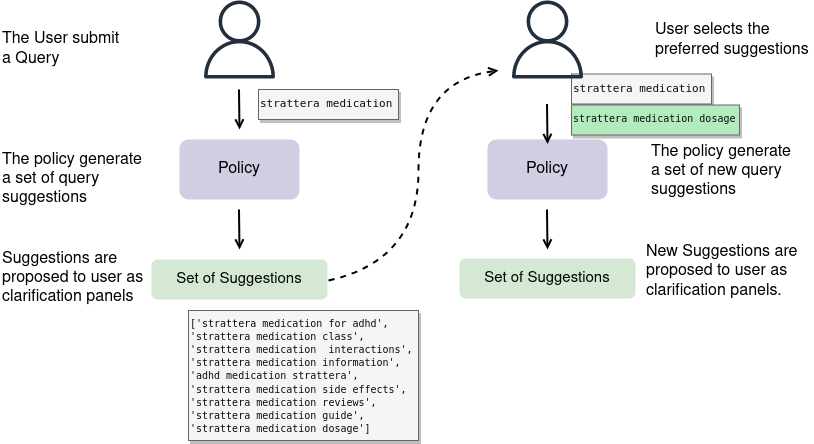}
    \caption{Multi-turn Query clarification framework}
    \label{fig:pipline}
\end{figure*}

\section{Related Work }

%In this section, we describe related works about interactive query clarification, Reinforcement Learning algorithm to fine-tune Language Model at sequence level, and language model for sets generation.
\subsection{Query Clarifications}

Interactive query clarification is a task that allows the user to be more involved in the query refinement process by interacting with the system \cite{10.1145/3020165.3020183}. Most modern search engines now assist the user with clarification panels or query suggestions in order to help refine and clarify their information need. While being referred to by different names in the literature, these suggestions serve the same purpose: to provide additional information to the user that may be relevant to their query. The format of both is also similar, typically appearing as a list of options for the user to choose from \cite{Zamani-Clarification,10.1145/3178876.3186068,10.1007/978-3-319-76941-7_54,DBLP:journals/corr/abs-1802-07997,Santos2012LearningTR,mustar:hal-02989015}.  This task can also be drawn as a multi-turn framework in which the user and the system interact iteratively \cite{10.1145/3020165.3020183,10.1145/3397271.3401467,abs-1909-12425, erbacher2022, Aliannejadi19-askingclarifying,10.1145/3397271.3401206}. In the pioneering work \cite{Aliannejadi19-askingclarifying}, the authors propose a conversation framework that consists in iteratively selecting a clarifying question at each conversation's turn. The main drawback of this approach, is that the conversation is defined a priori from (query/clarifying questions/user answers) triplets, crowdsourced and restrained to a limited number of topics. Zamani et al. \cite{Zamani-Clarification}  propose to use reformulation logs collected on real user to construct clarification panels. For a given query, they consider the most probable terms added by users when reformulating their queries. Additionally, they generate a clarifying questions associated to the topic using a LM. However, this model only consider one interaction with the user. Erbacher et al. \cite{erbacher2022} address the multi-turn clarification framework involving a user simulation to mimic user answer. However, suggestions are only generated based on the initial user query and are not updated between interaction turns.
This means that the maximal search performance is limited by the diversity in the pool.
In our work, we do regenerate iteratively the set of query clarifications based on previous user feedback. This means that we don't rely on a selection function as in \cite{erbacher2022} because our model can generate sets of reasonable size.

\subsection{Sequence level training}

Learning from users interaction is becoming a hot machine learning paradigm in IR. The community shows a growing interest for  Reinforcement Learning (RL) algorithms since IR systems deals with sequential interaction with users \cite{10.1145/3397271.3401467,10.1145/3397271.3401099,chandramohan:hal-00652446, chen2019generative, Nogueira2019MultiagentQR,abs-1909-12425,DBLP:journals/corr/NogueiraC17}. For instance, in \cite{Nogueira2019MultiagentQR,DBLP:journals/corr/NogueiraC17},  the authors proposed to use reinforcement learning for query expansion, the model was trained to maximize the expected  recall of relevant documents. In \cite{chen2019generative}, the authors used Reinforcement Learning to learn the user click policy.
In the context of language generation, RL can be used to adjust the model's output distribution to improve its performance on a specific task. One of the main difficulties in training a language model from scratch with RL is the large action space, this is why previous works used pretrained LM to constrain the exploration \cite{DBLP:journals/corr/BahdanauBXGLPCB16,DBLP:journals/corr/RanzatoCAZ15,DBLP:journals/corr/abs-2109-09371,DBLP:journals/corr/abs-1909-08593, Nogueira2019MultiagentQR,DBLP:journals/corr/NogueiraC17,Ouyang2022TrainingLM}. These methods are used for various NLP task with non-differentiable automatic reward function.  In \cite{DBLP:journals/corr/RanzatoCAZ15}, authors use automatic metrics BLEU \cite{10.3115/1073083.1073135} for translation task.  Some work also use human feedback to fine-tune their model \cite{Ouyang2022TrainingLM,DBLP:journals/corr/abs-1909-08593}. Authors  fine-tuned a large LM using Proximal Policy Optimization to generate better summarization according to a reward model trained on real human feedback.
Liu et al \cite{liu-etal-2020-learning} proposed to fine-tune their LM to generate more diverse paraphrase. However, in their work they use the ROUGE score between the generated sequences and a  reference sequence while in this work we use a fixed reference model as an anchor. In this paper we use Reinforcement Learning to fine-tune the LM to generate diverse sets of query suggestions. The model maximizes a dissimilarity metric while being grounded to a reference model.

\section{A generative model for multi-turn query clarification with  reinforcement learning}

\subsection{General overview}
Our model, called CIRCLE, aims at generating a set of query clarifications using reinforcement learning and is evaluated  within the sequential interaction framework proposed in \cite{Aliannejadi19-askingclarifying,erbacher2022}. This framework shown in the figure \ref{fig:pipline} consists in iteratively proposing query suggestions to the user in response to his/her query. The user feedback collected at each interaction step are used to generate the following set of suggestions. A retrieval model can be launched at each interaction to retrieve documents and evaluate the quality of the selected query. 
Formally, let consider the user query as a sequence of token $x =  w_{0},..., w_{t-1}$ sampled from a distribution of queries $D$, the goal is to learn a model using policy $\pi(w_t | w_{t-1},..., w_{0})$ that completes the sequence $x$ producing the set of queries $Y = Y = \{y_1, ...y_K\}$. Queries are separated with a $<sep>$ token in the sequence. 

With this in mind, our model is based on the following intuitions: 
\begin{itemize}
    \item Language models, well-known in the literature \cite{radford2019language}, have demonstrated their skills in generating sequences. We will therefore rely on decoders to generate query clarifications on the basis of an initial information need. 
    \item To provide choice to the user, we need to generate a set of query clarifications. We  leverage supervised pairs of (initial query-set of query suggestions) to fine-tune the model to generate set of query clarifications in an auto-regressive learning. In other words, decoder only architecture is exploited to generate a sequence of tokens that expresses the sequence of query clarifications, separated with a special token.
    \item To force the diversity of the different generated query clarifications within the sequence, we exploit reinforcement learning techniques estimating the similarity between clarifications while maintaining the distribution near the supervised model. 
\end{itemize}

%Our multi-turn query clarification model is adopted from \cite{Aliannejadi19-askingclarifying} [and Erbacher 2022], it enables multiple sequential interactions with a user to refine the query before retrieving any documents.
%More particularly, our framework is illustrated in Figure \ref{fig:pipline}. After issues an initial query $x$ associated to her/his information need $d$ at step $i=0$, at each time step $t$ the user iteratively select a one of the suggested query from the set $Y^i=\{y^i_1,y^i_2,...,y^i_K\}$ generated by the Language model. After $N$ turns, the IR system considers the last selected query and runs a retrieval model to retrieve documents. We assume that the number of proposed query suggestions should stay relatively low (between 2 and 10) because it may not be realist for a user to deal with large sets of suggestions.

\subsection{Query clarification Policy}
The objective of our CIRCLE policy is to generate a set of query clarifications which are both 1) grounded near the user query and 2) enough diverse to let the user explore the potential universe of keywords able to specify his/her need. To do so, our model is based on a reinforcement learning approach designed on the top of a supervised Language Model (LM). 
It is worth noting that the reinforcement learning approach is focused at the iteration level to generate more diverse sets of query suggestions for each turn, but does not embed a multi-turn level policy. %However, our model is fine-tuned using reinforcement learning to generate more diverse sets of query suggestions for each turn. The training process is described in the following section.

In what follows, we described these two components of our model.\\

%The policy in our framework is a pretrained LM that is finetuned for generating set of query suggestions. The model was initially trained, on a large corpus, to predict the next word, given previous words in some text\cite{ (Radford et al., 2019)} $P(w_t | w_{t-1},..., w_{0})$. Because it was pretrained on an agnostic task, the model is not optimized for a specific task. In this section we describe how we fine-tuned the LM to generate diverse sets of queries. (Following \cite{ziegler} we performed the experiment using a transformer architecture. We used the pretrained GPT2 model as our policy.) \\

\textbf{Supervised LM learning}

Numerous work heavily relies on LM and, particularly their decoding mechanism to generate sets of sequences \cite{VijayakumarCSSL16-DiverseBeamSearch, ,DBLP:journals/corr/abs-1810-05241,DBLP:journals/corr/abs-2009-10229,DBLP:journals/corr/abs-1904-09751,ye-etal-2021-one2set}. As LM are initially trained on a large corpus to predict the next word, given previous words in some text \cite{radford2019language} $P(w_t | w_{t-1},..., w_{0})$, they manage to learn common pattern and word associations in natural languages. In this work we propose to rely on a pretrained LM to fine-tune it to complete sequences of queries. Following \cite{DBLP:journals/corr/abs-1810-05241}, our objective is to fine-tune the pretrained LM to generate sequences of query suggestions  $Y = \{y_1, ...y_K\}$ given an input query $x$ using the cross-entropy loss. Query suggestions are separated with a $<sep>$ token in the sequence. In this work we use a decoder only architecture, this means that the sequence can be written: $x <sep>  y_1 <sep> ... y_K $

Despite various decoding tricks already used in the literature \cite{VijayakumarCSSL16-DiverseBeamSearch,DBLP:journals/corr/abs-1904-09751}, generating set of sequences using LM decoders suffer from the lack of diversity within the generated set as new decoded sequences are not conditioned on other previously decoded sequences. We present in what follows our strategy to  control the diversity of generated query clarification set by using reinforcement learning.\\ 

\textbf{Reinforcement Learning.}
 To improve the diversity of generated suggestions, the  language model is also fine-tuned  using the proximal policy optimization (PPO) \cite{schulman2017proximal} algorithm to maximize the expected reward $R$. Given a state $s_t = ( w_{t-1},..., w_{0})$ the policy predicts the next token $w_t$ maximizing the expected reward:
\begin{equation}
    \mathbb{E}_{\pi}(R) = \mathbb{E}_{x\sim D , Y\sim \pi(.|x) }[R(x,Y)]
     \label{maxexpr}
\end{equation}
With $R$ the reward function and $\pi$ the policy, $x$ the user query and $Y = \{y_1, ...y_K\}$  the set of query suggestions.

The value function (critic) used in PPO is initialized to the parameters of the reference model. However, the last linear layer is randomly initialized. For each state, this model predicts a value $V(s_t)$ used to compute an estimation of the advantage function $A = R(x,Y) - V(s_t)$. The critic is optimized to minimize the following loss function:
\begin{equation}
    L_v = \sum^T_{t=0} | R(x,Y) - V(s_t)|^2 
    \label{ppoloss}
\end{equation}

Using PPO the policy objective is written:
\begin{eqnarray}
    L_{\pi} &= & \mathbb{E}_{\pi} [min( A(w_t,s_t)\frac{\pi(w_t|s_t)}{\pi_{\text{old}(w_t|s_t)}}, \label{ppoloss2} \\ 
    &&A(w_t,s_t) clip(1- \epsilon, \frac{\pi(w_t|s_t)}{\pi_{\text{old}(w_t|s_t)}} , 1+\epsilon)) ] \nonumber
\end{eqnarray}
Note that, PPO smooths the policy $\pi$ update by constraining it to be close to the previous policy $\pi_{\text{old}}$ by clipping the gradients.

In this work, we want to improve the diversity in the set generated of our supervised model. For this purpose, the reward function used in this work is composed of 2 parts:
\begin{itemize}
    \item $r(x,Y)$ :Following \cite{liu-etal-2020-learning} we compute the dissimilarity between  between generated queries in the set:
\begin{equation}
    r = - \sum_{y_i \in \hat{Y}} \sum_{y_j \in \{ \hat{Y} \setminus y_i \}}  sim(y_i, y_j)
\end{equation}  
where $sim$ is a similarity metric computed between queries described in \ref{rbo} . 
\item The second part ensures that the policy stays near the supervised model. For this, we follow \cite{stiennon2020learning}, and add a KL term that penalizes the divergence between the policy $\pi$ and the reference model $\pi^{ref}$. This KL regularization makes sure the policy stays grounded near a fixed reference model. This guaranties that generated sequences by the policy are not far from generated sequences from the reference model. Otherwise, our model would likely generate diverse but random sequences.
The reward function can be written as:
\begin{equation}
    R(x,Y) = r(x,Y) - \beta \log[\frac{\pi^{RL}(Y|x)}{\pi^{ref}(Y|x)} ]
    \label{rewardfunction}
\end{equation}

\end{itemize}

\subsection{Model training and inference}
%We describe the training process in the following:
\begin{figure}[t]
    \centering
    \includegraphics[width=0.5\textwidth]{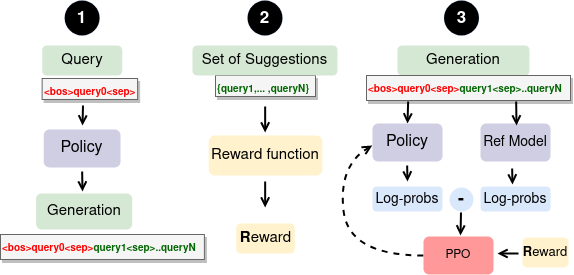}
    \caption{Fine-tuning GPT2 with PPO}
    \label{fig:ppo}
\end{figure}

The figure \ref{fig:ppo} shows the training of CIRCLE. The training is composed of 3 main steps:
\begin{enumerate}
\item Conditioned on the initial user query, the policy generates a trajectory, namely a sequence containing multiple queries. Queries are chained with  $<sep>$ token. We control the number of generated queries using stopping criteria counting the number of $<sep>$ token.
\item The reward is computed using the equation \ref{rewardfunction}.
\item The generated sequence is feed into the reference model and the policy. Resulting log-probabilities and reward are used to update the policy using PPO using the equations \ref{ppoloss} and \ref{ppoloss2} 
\end{enumerate}

%\subsection{Model evaluation}
During evaluation, the model is conditioned on the user query but also on the suggestions selected by the user at each interaction turn. This means that at each step $i$ the model complete the following sequence: $$ \text{input: }  <bos> x <sep>  y_{1}^+ <sep> ... y_{i}^+ $$ with  $q_{t=1}^+$ the selected query at step $i$ and $x$ the initial user query. 
%\subsection{Information retrieval model}
%The Information Retrieval (IR) model retrieves documents from a precomputed index. In the framework the query clarification is performed before any documents search and only rely on language modelling. Therefore, the IR model in this work is had-hoc and fixed, it takes as input a query and return a ranking of documents.

\section{Evaluation protocol}

The goal of this paper is to show that improving the diversity in the queries suggested to the user helps cover a wider range of topics and therefore better helps in the clarification process. To do so, the effectiveness of different models is evaluated in terms of how much they improve search performance in a multi-turn clarification framework. Therefore, we rely on classic IR metrics to evaluate these models. The user's actions are simulated based on various hypothesis.
%[faudra définir les objectifs d'évaluation : evaluer la strategie de reformulation via la taâche finale de RI. Est-ce qu'on évalue quelque chose sur les requêts ??? Faudra faire fitter les metriques avec ces objectifs]

\subsection{Dataset}

The experiment is conducted on MS Marco  2020 passages \cite{DBLP:journals/corr/NguyenRSGTMD16} which is an open domain datasets regrouping 8.8M passages and more than 500K Query-Passage relevance pairs. As  \cite{Nogueira2019MultiStageDR}, we trained our model on the train set of Msmacro and evaluate our model on a subset of the dev set (1200 queries sampled from 59 000).

\subsection{Metrics}
In accordance to the evaluation objectives, we rely on IR metrics to assess the effectiveness of the model in the multi-turn clarification framework but also to assess the diversity of proposed suggestions, we define different metrics. 

$\bullet$ To measure the quality of the selected query clarification, we evaluate the effectiveness of document ranking issued from this query using the well-known Mean Reciprocal Rank (MRR).
\begin{equation}
\text{MRR} = \frac{1}{|Y|} \sum^{|Y|}_i \frac{1}{\text{rank}_i}
\label{mrr}
\end{equation} 
with $\text{rank}_i$ the position of the first relevant document for the query $y_i \in Y$

$\bullet$ To measure the similarity between document rankings at different iterations, we rely on the Rank-Biased Overlap (RBO) \cite{10.1145/1852102.1852106}. This measures the similarity between incomplete and non-conjoint rankings and also values more heavily top ranked document. The more diverse the rankings are, the lower the score is.
\begin{equation}
\text{RBO}(S,T,p) =  (1-p)\sum^{\infty}_{d = 1} p^{d-1} . A_d
\label{rbo}
\end{equation}
With S, T are two document rankings, d is the actual depth of the ranking. $A_d$ expresses the agreement (the size of the intersection of both ranking) at depth $d$: $A_d = \frac{ | S_{:d} \cap T_{:d}|}{d}$ . $p$ determines the weight given the top ranked document.

\subsection{User Simulation}
Because real human feedback is time-consuming and costly, the information retrieval community often relies on user simulation to train or evaluate models \cite{DBLP:journals/corr/abs-2201-03435, Eckert1997UserEvaluation, Komatani2005UserGuidance, Pietquin2004A1, Schatzmann2006AStrategies, Scheffler2000ProbabilisticDialogues}.
We therefore design user simulation based on hypothesis that users does not always fully cooperate with IR systems. For instance a user may not know what suggestions might help. The goal is not to have realist behavior but to see the robustness and the limits of the proposed model. The proposed user simulation follows an epsilon greedy policy in which the user chooses a random action with a probability epsilon, and chooses the best known action with probability $1-\epsilon$. The $\epsilon$ parameter enables a wide range of different possible user behaviors from the most cooperative user to the most random user.  

It is worth noting that in our evaluation framework the user do not have stop criteria to end the search session. This allows us to observe all possible scenarios and record performance at every interaction turn. Moreover, we do not take into account the position bias that would be inducted by the position of the suggestions when presented to the user.

\subsection{Baselines}
We evaluated various methods to generate sets of query clarifications. These baselines are tested under different scenarios following the evaluation framework described in previous sections. \\
\textbf{Google suggestions:} Suggestions generated by Google search engine through their public API. Because google suggestions only rely on the previous considered query. We mimic multi-turn interaction by iteratively switching the previous query with the selected query by the user. Depending on the queries' specificity, between 1 and 10 google suggestions are proposed. We report the mean number of suggestions in the table \ref{table1}. The suggestions also depends on the language and the geographic location. In this work we use English for the United States. This can be considered as a strong baseline, thanks to its billions of users logs and feedback \footnote{How Google autocomplete works in Search: https://blog.google/products/search/how-google-autocomplete-works-search/}. However, we do acknowledge that the performance of these suggestions might be truncated because these suggestions are not specific for the Msmarco corpus and might be affected by current trends. \\
\textbf{Interact+Kmean:} This is the multi turn model proposed by \cite{erbacher2022}. We evaluated this model using the K-mean selection mechanism as suggested by the authors, selecting the best 2 ranked queries from different cluster in the pool of  64 queries.\\
\textbf{Beam search:} We fine-tuned a GPT2 model to perform one to one reformulations. We use beam search to generate a set of query suggestions.\\
\textbf{Supervised:} we use a GPT2 fine-tuned using supervised learning to complete sequences of queries. This model was trained on Google Suggestions collected. This baseline is the supervised version of our CIRCLE model without the reinforcement learning fine-tuning.\\
\textbf{CIRCLE:} The overall version of our  model, including the fine-tuning using Reinforcement Learning to leverage more diverse sets of query suggestions.\\

\subsection{Implementation details}
For the IR model, we opted for a Bert-base Dense retrieval model \cite{Hofstatter-topicaware}. This model was trained on Msmarco-passages to maximize the dot product between queries and their associated relevant passages. We used this model to compute documents embeddings. Embeddings are stored and indexed using Faiss HSWN32 index \cite{faiss}. We used the pretrained GPT2 provided by \cite{DBLP:journals/corr/abs-1910-03771}.
The same model is used to compute queries embeddings.
For supervised learning we used a learning rate of $lr = 2e^-5$  with batch size of 128. We trained the model on 3 epochs.
For finetuning with reinforcement learning we use a learning rate of $lr = 0.8e^-6$ with batch size of 128. The exploration is constrained to $top_p= 0.9$ and $top_k = 20$. We use $\beta = 0.01 $ and a clip ratio $\epsilon = 0.1$. The similarity metric used is the RBO \ref{rbo}. We use greedy decode the circle model using greedy decoding. We used $p=0.9$ in the RBO metrics. This means that the first ten documents are weighting for $85\%$ of the overall score.

\section{Results}
In this section, we report results that were assessed by following the evaluation protocol described above.

\subsection{Retrieval Performance}
In the table \ref{table1}, we report IR scores of different approaches to generate sets of suggestions on a subset of the devset of Msmsarco passages. We set the user fully cooperative ($\epsilon = 0$). Models are evaluated using the MRR metric (Equation \ref{mrr}).
\begin{table*}[t]
\centering
% Uncomment the line below to make the font size smaller if necessary
% \small
\adjustbox{max width=\textwidth}{%
\begin{tabular}{|l|c|c|c|c|c|c|c|}
\hline
 & i=0 (No interaction) & i=1 & i=2 & i=3 & i=4 & i=5 \\
\hline
User Query  & 0.2419 & - & - & - & - & - \\
\hline
Google Suggestions k = $\sim$(7.5/5.4/5.3/5.3/4.7)  & 0.2419 & \textbf{0.3820} & \textbf{0.3997} & \textbf{0.4034} & \textbf{0.4043} & \textbf{0.4049} \\
\hline
Interact+Kmean \cite{erbacher2022}/ k=2 (Kmean 64)  & 0.2419  &  0.1990 & 0.2195 & 0.2370 & 0.2478  & 0.2551\\
\hline
Beam Search k = 2  & 0.2419 & 0.2108 & 0.1872 & 0.1636 & 0.1536 & 0.1467 \\
Beam Search k = 4  & 0.2419 & 0.3169 & 0.2686 & 0.2573 & 0.2457 & 0.2434 \\
Beam Search k = 8  & 0.2419 & 0.3740 & 0.3469 &  0.3416 & 0.3295 &  0.3259 \\
\hline

Supervised k = 2  & 0.2419  & 0.2859 &  0.3017 &   0.3080 & 0.3074 &  0.3102 \\
Supervised k = 4  & 0.2419  & 0.3030 &  0.3298 &   0.3350 & 0.3357 &  0.3354 \\
Supervised k = 8  & 0.2419  & 0.3224 & 0.3614 &   0.3654 & 0.3649 &  0.3660 \\
\hline
CIRCLE k = 2  & 0.2419  &  0.2824 & 0.3063 & 0.3174 & 0.3277 & 0.3290   \\
CIRCLE k = 4  & 0.2419 & 0.3006 & 0.3250 & 0.3478 & 0.3614 & 0.3734 \\
CIRCLE k = 8   & 0.2419 & 0.3244 & 0.3555 & 0.3737 & 0.3844 & 0.3889 \\
\hline
\end{tabular}}
\caption{Effectiveness results on the subset of MS Marco passage 2020 dev set (1200 queries - 1 relevant document per query), k the size of the set of suggestions for interaction.}
\label{table1}
\end{table*}

\subsection{Suggestions Coverage}
In this section, we analyze the coverage of proposed suggestions models. To analyze this, we compute the mean similarity between ranking associated to each suggestion using the RBO metric (Equation \ref{rbo}).
 \begin{center} 
\begin{table*}[]
\resizebox{0.95\linewidth}{!}{
\begin{tabular}{l|l|l|l|l|}
\cline{2-5}
                          & Google Suggestions k = 5.64          & Beam search k=8             & Supervised k=8              & CIRCLE k=8                  \\ \hline
\multicolumn{1}{|l|}{Mean RBO} & \multicolumn{1}{c|}{0.3243} & \multicolumn{1}{c|}{0.3610} & \multicolumn{1}{c|}{0.3998} & \multicolumn{1}{c|}{0.3604} \\ \hline
\end{tabular}}
\caption{This table shows the mean ranking similarity between different model suggestions. The metric used is the RBO. The more diverse the rankings are, the lower the score is.}
\label{table2}
\vspace{-0.3cm}
\end{table*}

\end{center}

\subsection{Impact of user cooperation}
In this section we assess performance of the Google suggestions and CIRCLE interacting with non-perfect user. The objective is to observe the robustness of models with variable feedback quality. The user is following a greedy epsilon policy with epsilon ranging from $0.0$ to $0.5$.

\begin{figure}[!h]
    \centering
    \includegraphics[width=0.5\textwidth]{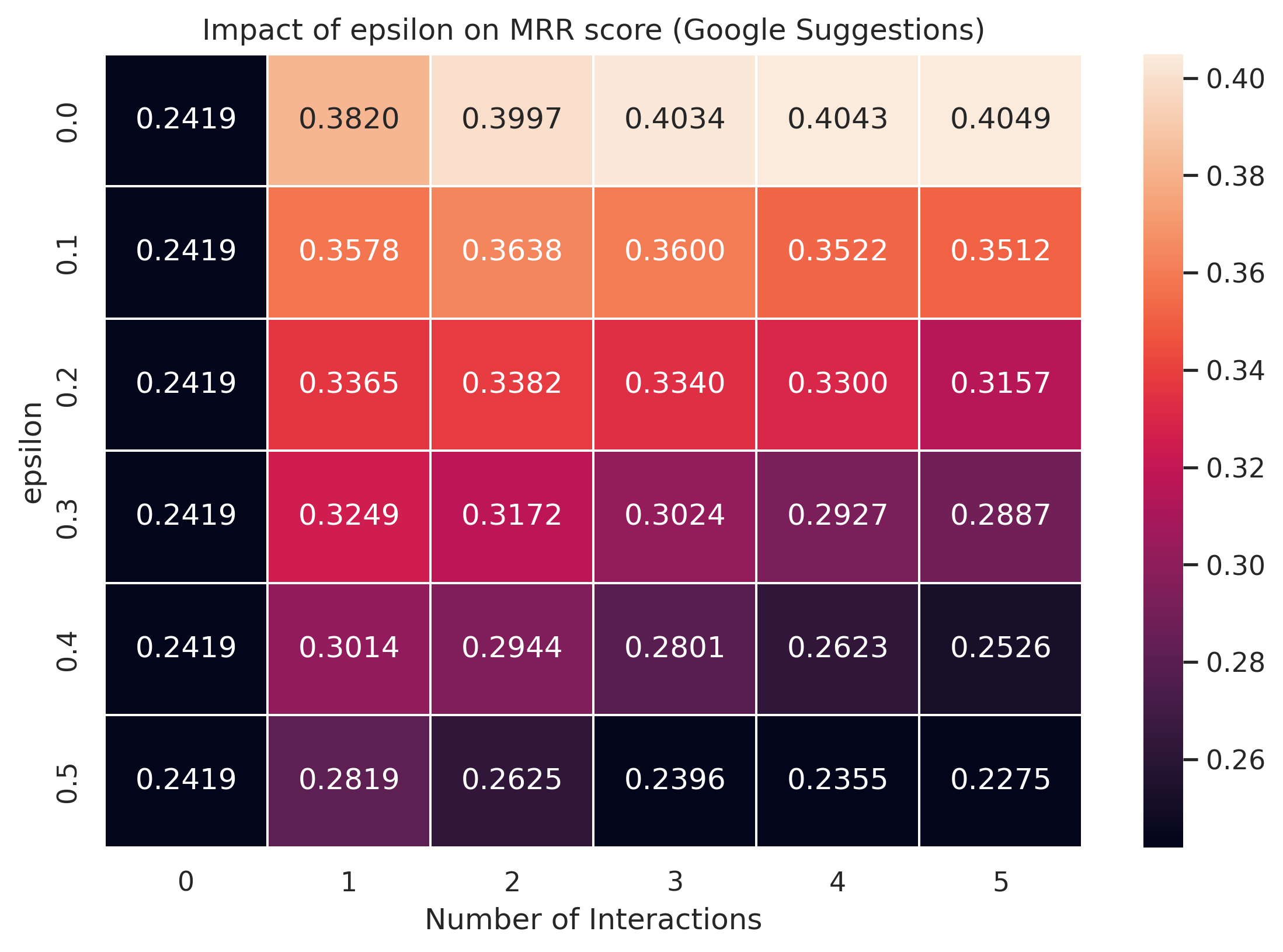}
    \caption{Google suggestions: Impact of the interactions with various user hypothesis on the system efficiency. The observed score is the MRR@1000.}
    \label{fig:heatgoogle}
\end{figure}
\vspace{-0.3cm}

 \begin{figure}[!h]
    \centering
    \includegraphics[width=0.5\textwidth]{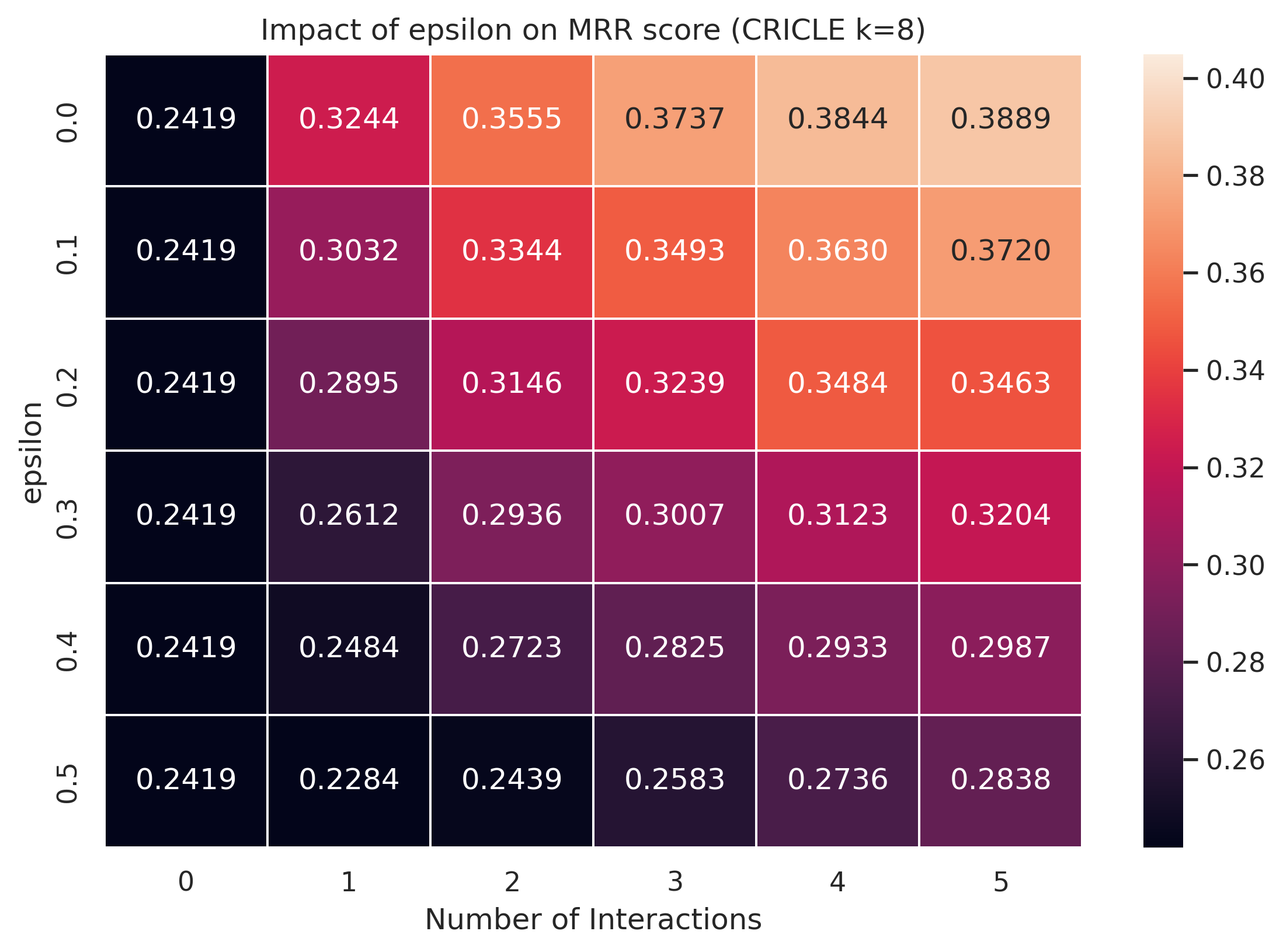}
    \caption{CIRCLE: Impact of the interactions with various user hypothesis on the system efficiency. The observed score is the MRR@1000.}
    \label{fig:heatcircle}
    \vspace{-0.5cm}

\end{figure}

\section{Results Analysis}
In the table \ref{table1}, we see that Google suggestions are a strong baseline for this task. With only an average of $5.64$ query suggestions it manages to reach a MRR of $0.4049$. For $k=8$ , the beam search manage to reach a MRR of $0.3740$ in the first interaction. However, the score decreased with each subsequent interaction. This can be explained because the model takes as input only the last selected query and therefore the model generation can diverge if generated queries are less efficient than in the previous turn and because the user always select one of the generated queries. CIRCLE  manage to improve the mean score from $0.3244$ in the first  interaction to $0.3889$ in the last interaction. Against $0.3224$ to $0.3660$ for the supervised model. This means that the additional RL fine-tuning was beneficial to improve the overall performance.
The table \ref{table2} shows that Google suggestions have a good coverage of possible user needs with a mean RBO of $0.3243$ while proposing only an average of $5.64$ query suggestions. Surprisingly, the beam search also has a good coverage with a RBO of $0.3610$. We can see that the RL fintuning improve the mean RBO between the supervised and CIRCLE. CIRCLE and the beam search have sensibly the same RBO.
We can see in the figures \ref{fig:heatgoogle} and \ref{fig:heatcircle} that none of the proposed model is robust to non-cooperative user. We observe that our model is abit more resilient with a non fully-cooperative user. This is explained because our model takes as input all the sequence of previously selected queries including the initial user query. Whereas google suggestions and beam search only consider the last selected query as input. This means that when the user select a query not-relevant for his/her information need, the new generated suggestions diverges from the initial user query.
\section{Conclusion and perspectives}
In this work, we consider the multi-turn interactive query clarification problem for information retrieval. We propose CICRLE, a gpt2 model that generates a sequence of queries suggestions conditionally to one or several queries. By using reinforcement learning, we manage to increase the set diversity compared to the supervised model. Additionally, this model completes sequences composed of past feedback (selected queries) to generate new sets of queries iteratively, showing the benefit of multi-turn clarification. We especially when with non non-perfect user, relative to google suggestions. However, there are several downsides: the proposed model is far from Google suggestions performances, and it requires several interactions with the user to perform better than the beam search. Additionally, the proposal model does not embed strategy at interaction level, and was not optimized on any user behavior or IR metrics.    There are several ways this work can be extended.  1)The policy can be trained to maximize a utility function with multi turn interaction with the user. The quality of interaction between user/system could benefit strongly. However, the main difficulties are that the rewards are very sparse, and incorporating interaction feedback in the model is challenging during the training.
2)The performance of the model may be improved by using a loss invariant to queries permutation during supervised training. Note that we tried to reproduce the method in  \cite{ye-etal-2021-one2set} and apply it with a pretrained LM on query suggestions task. However, we were not able to control the language model generation with the control code without deteriorating the generation quality. 3)The model can be extended to leverage search session or multi-facets information needs. Generated suggestions based on multiple user information needs and therefore multiple queries requiring user simulation  to mimic search session \cite{DBLP:journals/corr/abs-2201-11181}.

%\subsection{References}

%\section*{Acknowledgements}

% Entries for the entire Anthology, followed by custom entries
\bibliography{anthology,custom}
\bibliographystyle{acl_natbib}

\appendix

\end{document}